\documentstyle[12pt]{article}
\newlength{\mathspace}
\def\np#1{ Nucl. Phys. B#1}
\def\pr#1    { Phys. Rev. D#1 }
\def\pl#1{ Phys. Lett. B#1}
\def\cmp   { Commun. Math. Phys. }

\def\ijmp#1  { Int. Jour. Mod. Phys. A#1 }
\def\mpl#1   { Mod. Phys. Lett. A#1 }
\def\begineq{\begin{equation}}
\def\endeq{\end{equation}}
\def\eqabegin{\begin{eqnarray}}
\def\eqaend{\end{eqnarray}}
\def\nn{\nonumber}

\def\parmedskip        {  \par\medskip  }

\def\parbigskipn        {  \par\bigskip\noindent  }

\begin{document}
\baselineskip=0.7cm
\setlength{\mathspace}{2.5mm}



\begin{titlepage}

    \begin{normalsize}
     \begin{flushright}
                 UR-1445\\
                 US-FT-28/95 \\
                 hep-th/9511091 \\
     \end{flushright}
    \end{normalsize}
    \begin{LARGE}
       \vspace{1cm}
       \begin{center}
         {The Zero Curvature Formulation of }\\
         {TB, sTB Hierarchy and Topological Algebras} \\
       \end{center}
    \end{LARGE}

  \vspace{5mm}

\begin{center}
           Ashok D{\sc as}

           \vspace{2mm}

             {\it Department of Physics and Astronomy}\\
             {\it University of Rochester}\\
             {\it Rochester, N.Y. 14627, USA}\\

           \vspace{.5cm}

                \ \  and \ \

           \vspace{.5cm}

           Shibaji R{\sc oy}
           \footnote{E-mail address:
              roy@gaes.usc.es}

                 \vspace{2mm}

        {\it Departamento de F\'\i sica de Part\'\i culas} \\
        {\it Universidade de Santiago}\\
        {\it E-15706 Santiago de Compostela, Spain}\\
      \vspace{1cm}

    \begin{large} ABSTRACT \end{large}
        \par
\end{center}
 \begin{normalsize}
\ \ \ \
A particular dispersive generalization of long water wave equation in $1+1$
dimensions, which is  important in the study of matrix models without
scaling limit, known as two--Boson (TB) equation, as well as the associated
hierarchy has been derived from the zero curvature condition on the gauge
group $SL(2,R)\otimes U(1)$. The supersymmetric extension of the two--Boson
(sTB) hierarchy has similarly been derived from the zero curvature condition
associated with the gauge supergroup $OSp(2|2)$. Topological algebras arise
naturally as the second Hamiltonian structure of these classical integrable
systems, indicating a close relationship of these models with 2d topological
field theories.
\end{normalsize}

\end{titlepage}
\vfil\eject
Integrable models [1--4] in $1+1$ dimensions play very interesting
and sometimes
mysterious roles in the study of conformal field theories [5], matrix model
formulation of string theories [6], 2d topological field theories [7,8]
and the
intersection theory on the moduli space of Riemann surfaces [9,10].
These models
contain very rich mathematical structures and deserve to be studied on their
own right. Recently, a particular dispersive generalization of long water
wave equation in narrow channel, known as the TB equation [11] has
attracted some
attention due to its relation with the matrix models without the continuum
limit [12]. This relation has been interpreted as an indication of the
topological nature of the latter models.

The TB--hierarchy as well as their close relatives have been studied quite
extensively in the literature [11,13,14]. In particular, TB--hierarchy
was constructed
in ref.[11] in a non--standard Lax operator approach and has been shown to
possess a tri--Hamiltonian structure. The supersymmetric extension of
TB--hierarchy or sTB--hierarchy and many of their interesting properties have
also been studied recently [15,16] in the non--standard superLax
operator approach.
Integrable models, on the otherhand, are also known [17] to be obtained from
the group theoretical point of view where the dynamical equations are obtained
from the zero curvature condition associated with some symmetry group. While
some properties of an integrable model can be understood in the Lax operator
approach, the bi--Hamiltonian structure, integrabilty property and the geometry
of the phase--space become more transparent in the zero curvature formulation
[18].
Moreover, ordinary 2d gravity theories and their underlying current algebra
structure in the light cone gauge can be better understood by connecting them
to integrable models in the zero curvature approach [19,20].
Thus, these two approaches play complementary roles with each other.

In this paper we study both TB and sTB--hierarchy in the zero curvature
approach. We derive TB--hierarchy from the zero curvature condition associated
with the gauge group $SL(2,R)\otimes U(1)$ and similarly sTB--hierarchy is
obtained from the gauge supergroup $OSp(2|2)$. We point out that both these
integrable systems possess two bi--Hamiltonian structures. The second
Hamiltonian structure of the TB--hierarchy is shown to be $U(1)$ extended
Virasoro algebra with zero central charge and that of sTB--hierarchy is the
twisted $N=2$ superconformal algebra. This observation, therefore, indicates
that these integrable models are closely related to 2d topological field
theories.

The dispersive long water wave equation (TB equation) with which we will be
concerned in this paper has the form:
\eqabegin
\frac{\partial J_1}{\partial t} &=& \left(J_1' - 2 J_1J_0\right)'\nn\\
\frac{\partial J_0}{\partial t} &=& \left(2 J_1 - J_0' - J_0^2\right)'
\eqaend
These are coupled differential equations where $J_1(x,t)$ and $J_0(x,t)$ refer
respectively the horizontal velocity and the height of the free water surface
and a prime denotes derivative with respect to $x$. By choosing the scaling
dimension of $x$ to be $-1$, we find from eq.(1) that the scaling dimensions
of $t$, $J_1(x,t)$ and $J_0(x,t)$ are fixed as $-2$, 2 and 1 respectively. This
observation will be helpful in choosing the gauge fixing condition in our zero
curvature analysis.

Now we show how the TB equation (1) as well as the hierarchy associated with it
can be obtained from the zero curvature condition on a special gauge fixed form
of $SL(2,R)\otimes U(1)$ Lie algebra valued gauge fields. There are four
bosonic generators denoted by $t_a$ in $SL(2,R)\otimes U(1)$ Lie algebra, where
$a\,=\,0,\,\pm$ are the $SL(2,R)$ indices and $a\,=\,(0)$ is the $U(1)$ index.
The algebra can be written as,
\begineq
\left[t_a,\,t_b\right]\,\,=\,\,f_{ab}^c\,t_c
\endeq
where the non-zero structure constants are $f_{+0}^+\,=\, f_{0-}^-\,=\,\frac
{1}{2}f_{+-}^0\,=\,1$. The zero curvature condition in $1+1$ dimensions
associated with the Lie algebra valued gauge fields $A_\mu \equiv A_\mu^a t_a$,
where $\mu = t,\,x$ denote respectively the time and space components, has the
form:
\begineq
F_{tx}\,\,=\,\,\partial_t A_x - \partial_x A_t + \left[A_t,\,A_x\right]\,\,
=\,\,0.
\endeq
Written in components we get four equations from (3) as given below,
\eqabegin
\partial_t A_x^+ - \partial_x A_t^+ - A_t^0 A_x^+ + A_t^+ A_x^0 &=& 0\\
\partial_t A_x^- - \partial_x A_t^- + A_t^0 A_x^- - A_t^- A_x^0 &=& 0\\
\partial_t A_x^0 - \partial_x A_t^0 + 2 A_t^+ A_x^- - 2 A_t^- A_x^+ &=& 0\\
\partial_t A_x^{(0)} - \partial_x A_t^{(0)} &=& 0
\eqaend
Identifying $A_x^+ \equiv J_1(x,t)$ and $A_x^{(0)} \equiv J_0(x,t)$ as the two
dynamical variables of the TB equation (1), and noting that they have scaling
dimensions 2 and 1 respectively as mentioned before, we get from (4)--(7) the
scaling dimensions of the other two components of the gauge field as $\left[
A_x^-\right] = 0$ and $\left[A_x^0\right] = 1$. We, therefore, choose the
gauge fixing conditions as,
\begineq
A_x^+\,\,=\,\,J_1(x,t),\quad A_x^-\,\,=\,\,-1,\quad A_x^0\,\,=\,\,J_0(x,t),
\quad A_x^{(0)}\,\,=\,\,J_0(x,t)
\endeq
With these gauge fixing conditions, we note that from eq.(5) and from a
combination of (6) and (7) we get two constraint equations which can be
solved as,
\eqabegin
A_t^0 &=& -\partial_x A_t^- -J_0 A_t^-\nn\\
A_t^+ &=& \frac{1}{2}\partial_x A_t^{(0)} + \frac{1}{2}\partial_x^2 A_t^-
+ \frac{1}{2} J_0' A_t^- + \frac{1}{2}J_0 \partial_x A_t^- - J_1 A_t^-
\eqaend
Substituting these in (4) and (7), we obtain two dynamical equations as
follows,
\eqabegin
\partial_t J_1 &=& \frac{1}{2}\partial_x^2 A_t^{(0)} +
\frac{1}{2}\partial_x^3 A_t^- + \frac{1}{2} J_0'' A_t^-
+ J_0' \partial_x A_t^- - J_1' A_t^-\nn\\
& & - 2 J_1\partial_x A_t^- - \frac{1}{2}J_0\partial_x A_t^{(0)}
- \frac{1}{2} J_0J_0' A_t^- - \frac{1}{2}J_0^2 \partial_x A_t^-\nn\\
\partial_tJ_0 &=& \partial_x A_0^{(0)}
\eqaend
In (10) we have obtained the dynamical equations in terms of two independent
functions $A_t^-(J_1,J_0)$ and $A_t^{(0)}(J_1,J_0)$. We now make a redefinition
of the function $A_t^{(0)}$ in terms of the old variables $A_t^{(0)}$ and
$A_t^-$ as given below,
\begineq
{\tilde A}_t^{(0)}\,\,=\,\,\frac{1}{2} A_t^{(0)} + \frac{1}{2}\partial_x A_t^-
+ \frac{1}{2}J_0 A_t^-
\endeq
With this redefinition the dynamical equations (10) reduce to the following
form:
\eqabegin
\partial_t J_1 &=& \partial_x^2 A_t^{(0)} - J_0 \partial_x A_t^{(0)}
- J_1' A_t^- - 2 J_1 \partial_x A_t^-\\
\partial_t J_0 &=& 2\partial_x A_t^{(0)} - \partial^2 A_t^- -J_0' A_t^-
- J_0 \partial_x A_t^-
\eqaend
where in writing (12) and (13), we have omitted `tilde' from $A_t^{(0)}$. These
equations define the TB--hierarchy. We now show how to extract the explicit
forms of the equations of the hierarchy as well as the two bi-Hamiltonian
structures mentioned earlier from these hierarchy equations.

By shifting $J_0(x,t)\rightarrow J_0(x,t) + \lambda$, where $\lambda$ is a
space--time independent parameter of scaling dimension 1, known as the
spectral parameter, we obtain from the hierarchy equations (12) and (13),
\eqabegin
\partial_t J_1 &=& \partial_x^2 A_t^{(0)} - J_0 \partial_x A_t^{(0)}
- J_1' A_t^- - 2 J_1 \partial_x A_t^- -\lambda\partial_x A_t^{(0)}\nn\\
\partial_t J_0 &=& 2\partial_x A_t^{(0)} - \partial^2 A_t^- -J_0' A_t^-
- J_0 \partial_x A_t^- -\lambda\partial_x A_t^-
\eqaend
Since the dynamical variables $J_1(x,t)$ and $J_0(x,t)$ are independent of
$\lambda$, we obtain a set of recursion relation and dynamical equation from
(14) if we expand
\eqabegin
A_t^-(J_1,J_0,\lambda) &=& \sum_{j=0}^n\,A_j(J_1,J_0)\lambda^{n-j}\nn\\
A_t^{(0)}(J_1,J_0,\lambda) &=& \sum_{j=0}^n\,B_j(J_1,J_0)\lambda^{n-j}\nn
\eqaend
as follows,
\eqabegin
-\left(J_1' + 2 J_1 \partial_x\right)A_j + \left(\partial_x^2    -
J_0\partial_x\right)B_j &=& \partial_x B_{j+1}\nn\\
-\left(\partial_x^2 + J_0' +J_0\partial_x\right)A_j +
2\partial_x B_j &=& \partial_x A_{j+1}\qquad {\rm for}\,\, j=0,1,\ldots,n-1
\eqaend
and for $j=n$
\eqabegin
\partial_t J_1 &=& -\left(J_1' + 2 J_1 \partial_x\right)A_n +
\left(\partial_x^2 -
J_0\partial_x\right)B_n\nn\\
\partial_t J_0 &=& -\left(\partial_x^2 + J_0' +J_0\partial_x\right)A_n +
2\partial_x B_n
\eqaend
{}From these two sets of equations (15) and (16), we can generate the whole
hierarchy associated with TB equation as follows. It is clear from (15) that
$A_0$ and $B_0$ are both constants and therefore we choose
$A_0\,=\,B_0\,=\,-1$. Substituting these values in (16) we obtain the zeroth
order dynamical equation as given below,
\eqabegin
\partial_t J_1 &=& J_1'\nn\\
\partial_t J_0 &=& J_0'
\eqaend
which are nothing but the chiral wave equations. Then using the values of
$A_0$ and $B_0$ in the recursion relations (15) we obtain $A_1\,=\,J_0$ and
$B_1\,=\,J_1$. Substituting these values in (16) we get the first order
dynamical equations as,
\eqabegin
\partial_t J_1 &=& \left(J_1' - 2 J_1 J_0\right)'\nn\\
\partial_t J_0 &=& \left(2 J_1 - J_0' - J_0^2\right)'
\eqaend
These are precisely the TB equation written earlier in (1). This procedure,
therefore, generates the TB--hierarchy. By identifying $A_n\,\equiv\,\frac
{\delta H_n}{\delta J_1}$ and $B_n\,\equiv\,\frac{\delta H_n}{\delta J_0}$,
where $\delta$ indicates the variational derivatives and $H_n$'s are the
Hamiltonians (conserved quantities) associated with the integrable system,
we can easily recover from (15) and (16) the two Hamiltonian structures
(Poisson brackets) among the dynamical variables as,
\eqabegin
\{J_1(x),\,J_0(y)\}_1 &=& \partial_x \delta (x-y)\nn\\
\{J_1(x),\,J_1(y)\}_1 &=& \{J_0(x),\,J_0(y)\}_1\,\,\,=\,\,\,0
\eqaend
and
\eqabegin
\{J_1(x),\,J_1(y)\}_2 &=& -\left[J_1'(x) + 2 J_1(x)\partial_x\right]
\delta (x-y)\nn\\
\{J_1(x),\,J_0(y)\}_2 &=& \left[\partial_x^2 - J_0(x)\partial_x\right]
\delta(x-y)\nn\\
\{J_0(x),\,J_0(y)\}_2 &=& 2\partial_x\delta(x-y)
\eqaend
Using the Poisson bracket structures, one can easily work out the explicit
forms of the Hamiltonians $H_n$'s, in order to write the dynamical equations
as Hamilton's equation of motion. This way one can regard the integrable
systems as Hamiltonian systems. In order to extract the second bi-Hamiltonian
system, we can proceed exactly as in the previous case except that we now
shift instead, the other dynamical variable $J_1(x,t)$ by the
spectral parameter
as, $J_1(x,t)\rightarrow J_1(x,t) + \lambda^2$. Now expanding the independent
functions in terms of the spectral paramter as
\eqabegin
A_t^-(J_1,J_0,\lambda^2) &=&
\sum_{j=0}^n\,A_j(J_1,J_0)(\lambda^2)^{n-j}\nn\\
A_t^{(0)}
(J_1,J_0,\lambda^2) &=& \sum_{j=0}^n\,B_j(J_1,J_0)(\lambda^2)^{n-j}\nn
\eqaend
we
obtain the recursion relation in the following form
\eqabegin
-\left(J_1' + 2 J_1 \partial_x\right)A_j + \left(\partial_x^2 -
J_0\partial_x\right)B_j &=& 2\partial_x A_{j+1}\nn\\
-\left(\partial_x^2 + J_0' +J_0\partial_x\right)A_j +
2\partial_x B_j &=& 0\quad\qquad {\rm for}\,\, j=0,1,\ldots,n-1
\eqaend
and for $j=n$, the dynamical equations remain the same as eq.(16). It is
clear from (21) that in this case $A_0$ is a constant and $B_n$'s get fixed
from the second equation of (21). Choosing $A_0\,=\,-1$, we get the zeroth
and first order equations as,
\eqabegin
\partial_t J_1 &=& \left(-\frac{1}{2}J_0' + \frac{1}{4}J_0^2 + J_1\right)'\nn\\
\partial_t J_0 &=& 0
\eqaend
and
\eqabegin
\partial_t J_1 &=& -\frac{1}{8}J_0'''' + \frac{3}{4}J_0(J_0')^2
+\frac{3}{8}J_0^2 J_0'' + J_1 J_0'' + \frac{5}{4} J_1' J_0' + \frac{1}{8}
J_0J_0'''\nn\\
& & - \frac{3}{8}J_0^3J_0' + J_1J_0J_0' - \frac{5}{8} J_1'J_0^2 - \frac{3}{2}
J_1 J_1' + \frac{1}{2} J_1'''\nn\\
\partial_t J_0 &=& 0
\eqaend
We note that in this case one of the dynamical variables $J_0(x,t)$ does not
evolve with time. So, setting it to zero, we recover from (23) the KdV
equation. This shows how KdV--hierarchy is contained in TB--hierarchy. The
first Hamiltonian structure can be seen from (21) to have the form:
\eqabegin
\{J_1(x),\,J_1(y)\}_1 &=& 2\partial_x \delta (x-y)\nn\\
\{J_1(x),\,J_0(y)\}_1 &=& \{J_0(x),\,J_0(y)\}_1\,\,\,=\,\,\,0
\eqaend
and the second Hamiltonian structure remains as given before in (20). This
analysis, therefore, brings out the two bi--Hamiltonian structures associated
with the TB--hierarchy. We can recognize the topological nature of the second
Hamiltonian structure by observing that (20) represents a $U(1)$ extended
Virasoro algebra with zero central charge where the $U(1)$ current is
anomalous. This topological symmetry will become more manifest when we go to
the supersymmetric case i.e. sTB--hierarchy which we consider next.

We will show that sTB--hierarchy can be generated from the zero curvature
condition on the gauge supergroup $OSp(2|2)$ which consists of four bosonic
and four fermionic generators. The corresponding Lie superalgebra is given as
\begineq
\left[t_i,\,t_j\right]\,\,=\,\,f_{ij}^k t_k;\quad \left[t_i,\,t_\alpha\right]
\,\,=\,\,f_{i\alpha}^\beta t_\beta;\quad \left[t_\alpha,\,t_\beta\right]_+
\,\,=\,\,f_{\alpha\beta}^i t_i;\quad
\endeq
where the subscript `$+$' denotes the anticommutator. The bosonic indices
$i$, $j$ take values 0, $\pm$, $(0)$ and the fermionic indices $\alpha$,
$\beta$ take $\pm\frac{1}{2}$, $(\pm\frac{1}{2})$. The non--zero structure
constants are $f_{+0}^+\,\,=\,\,f_{0-}^-\,\,=\,\,f_{(0)\frac{1}{2}}
^\frac{1}{2}\,\,=\,\,f_{(0)-\frac{1}{2}}^{-\frac{1}{2}}\,\,=\,\,
f_{\frac{1}{2}(0)}^{(\frac{1}{2})}\,\,=\,\,f_{(-\frac{1}{2})(0)}
^{(-\frac{1}{2})}\,\,=\,\,f_{\frac{1}{2}(-\frac{1}{2})}^0\,\,=\,\,
f_{-\frac{1}{2}(\frac{1}{2})}^0\,\,=\,\,f_{-\frac{1}{2}(-\frac{1}{2})}^-
\,\,=\,\,f_{\frac{1}{2}(\frac{1}{2})}^+\,\,=\,\,f_{\frac{1}{2}-}
^{-\frac{1}{2}}\,\,=\,\,f_{+ -\frac{1}{2}}^\frac{1}{2}\,\,=\,\,
f_{+(-\frac{1}{2})}^{(\frac{1}{2})}\,\,=\,\,f_{\frac{1}{2} -}
^{(-\frac{1}{2})}\,\,=\,\,2 f_{\frac{1}{2}(-\frac{1}{2})}^{(0)}\,\,=\,\,
2 f_{\frac{1}{2} 0}^\frac{1}{2}\,\,=\,\, 2 f_{0 -\frac{1}{2}}
^{-\frac{1}{2}}\,\,=\,\, 2 f_{(\frac{1}{2}) 0}^{(\frac{1}{2})}\,\,=\,\,
2 f_{0 (-\frac{1}{2})}^{(-\frac{1}{2})}\,\,=\,\,-2 f_{(\frac{1}{2})
-\frac{1}{2}}^0\,\,=\,\,\frac{1}{2} f_{+ -}^0\,\,=\,\,1$. So, the zero
curvature condition (3) in this case in components takes the form:
\eqabegin
\partial_t A_x^+ - \partial_x A_t^+ + A_t^+ A_x^0 - A_t^0 A_x^+ + A_t^{\frac
{1}{2}} A_x^{(\frac{1}{2})} + A_t^{(\frac{1}{2})} A_x^{\frac{1}{2}} &=& 0\\
\partial_t A_x^- - \partial_x A_t^- + A_t^0 A_x^- - A_t^- A_x^0 + A_t^{-\frac
{1}{2}} A_x^{(-\frac{1}{2})} + A_t^{(-\frac{1}{2})} A_x^{-\frac{1}{2}} &=& 0\\
\partial_t A_x^0 - \partial_x A_t^0 + A_t^{\frac{1}{2}} A_x^{(-\frac{1}{2})}
+ A_t^{(-\frac{1}{2})} A_x^{\frac{1}{2}} + A_t^{-\frac
{1}{2}} A_x^{(\frac{1}{2})} + A_t^{(\frac{1}{2})} A_x^{-\frac{1}{2}} & &\nn\\
+ 2 A_t^+ A_x^- - 2 A_t^- A_x^+ &=& 0\\
\partial_t A_x^{(0)} - \partial_x A_t^{(0)} -\frac{1}{2} A_t^{(\frac{1}{2})}
A_x^{-\frac{1}{2}} - \frac{1}{2} A_t^{-\frac{1}{2}} A_x^{(\frac{1}{2})} +
\frac{1}{2} A_t^{\frac{1}{2}} A_x^{(-\frac{1}{2})} + \frac{1}{2} A_t
^{(-\frac{1}{2})} A_x^{\frac{1}{2}} &=& 0\\
\partial_t A_x^{\frac{1}{2}} - \partial_x A_t^{\frac{1}{2}} + A_t^{(0)}
A_x^{\frac{1}{2}} - A_t^{\frac{1}{2}} A_x^{(0)} + A_t^+ A_x^{-\frac{1}{2}} -
A_t^{-\frac{1}{2}} A_x^+ & &\nn\\
+ \frac{1}{2} A_t^{\frac{1}{2}} A_x^0 -
\frac{1}{2} A_t^0 A_x^{\frac{1}{2}} &=& 0\\
\partial_t A_x^{-\frac{1}{2}} - \partial_x A_t^{-\frac{1}{2}} + A_t
^{\frac{1}{2}} A_x^- - A_t^- A_x^{\frac{1}{2}} + A_t^{(0)} A_x^{-\frac{1}{2}}
- A_t^{-\frac{1}{2}} A_x^{(0)} & &\nn\\
+ \frac{1}{2} A_t^0 A_x^{-\frac{1}{2}} -
\frac{1}{2} A_t^{-\frac{1}{2}} A_x^0 &=& 0\\
\partial_t A_x^{(\frac{1}{2})} - \partial_x A_t^{(\frac{1}{2})} + A_t^+
A_x^{(-\frac{1}{2})} - A_t^{(-\frac{1}{2})} A_x^+ + A_t^{(\frac{1}{2})}
A_x^{(0)} - A_t^{(0)} A_x^{(\frac{1}{2})} & &\nn\\
+ \frac{1}{2} A_t^{(\frac{1}{2})}
A_x^0 - \frac{1}{2} A_t^0 A_x^{(\frac{1}{2})} &=& 0\\
\partial_t A_x^{(-\frac{1}{2})} - \partial_x A_t^{(-\frac{1}{2})} + A_t
^{(\frac{1}{2})} A_x^- - A_t^- A_x^{(\frac{1}{2})} + A_t^{(-\frac{1}{2})}
A_x^{(0)} - A_t^{(0)} A_x^{(-\frac{1}{2})} & &\nn\\
+ \frac{1}{2} A_t^0 A_x
^{(-\frac{1}{2})} - \frac{1}{2} A_t^{(-\frac{1}{2})} A_x^0 &=& 0
\eqaend
Identifying $A_x^+\,\equiv\,J_1(x,t)$, $A_x^{(0)}\,\equiv\,\frac{1}{2}J_0
(x,t)$ as the
bosonic variables of sTB--hierarchy and $A_x^{(\frac{1}{2})}\,\equiv\,\xi
(x,t)$,
$A_x^{\frac{1}{2}}\,\equiv\,\bar {\xi}(x,t)$ as the fermionic variables, a
simple
dimensional analysis from (26)--(33) allows us to choose the following gauge
fixing conditions for the rest of the gauge field components,
\begineq
A_x^-\,\,=\,\,-1, \quad A_x^0\,\,=\,\,J_0(x,t), \quad A_x^{-\frac{1}{2}}
\,\,=\,\,A_x^{(-\frac{1}{2})}\,\,=\,\,0
\endeq
Substituting these values, we find that eqs.(27), (31), (33) and a
combination of (28) and (29) give four constraints which can be solved to
obtain,
\eqabegin
A_t^0 &=& -\partial_x A_t^- - A_t^- J_0\nn\\
A_t^{\frac{1}{2}} &=& -\partial_x A_t^{-\frac{1}{2}} - A_t^- \bar {\xi}
- A_t^{-\frac{1}{2}}J_0\nn\\
A_t^{(\frac{1}{2})} &=& -\partial_x A_t^{(-\frac{1}{2})} - A_t^- \xi\\
A_t^+ &=& \frac{1}{2} \partial_x A_t^{(0)} + \frac{1}{2} \partial_x^2 A_x^-
+ A_t^{-\frac{1}{2}}\xi + \frac{1}{2}\partial_x A_t^- J_0 + \frac{1}{2}
A_t^- J_0' - A_t^- J_1\nn
\eqaend
and the other equations (26), (29), (30) and (32) give the dynamical equations
in the following form
\eqabegin
\partial_t J_1 &=& -\left(J_1' + 2 J_1 \partial_x\right) A_t^- +
\left(\partial_x^2
- J_0 \partial_x\right) A_t^{(0)} - \left(\xi' + 2 \xi \partial_x\right)
A_t^{-\frac{1}{2}} - \bar {\xi} \partial_x A_t^{(-\frac{1}{2})}\nn\\
\partial_t J_0 &=& -\left(\partial_x^2 + J_0 \partial_x + J_0'\right) A_t^-
+ 2 \partial_x A_t^{(0)} - \xi A_t^{-\frac{1}{2}} + \bar {\xi}
A_t^{(-\frac{1}{2})}\nn\\
\partial_t \xi &=& -\left(2 \xi \partial_x + \xi'\right) A_t^- + \xi A_t^{(0)}
- \left(\partial_x^2 - J_1 - J_0\partial_x\right) A_t^{(-\frac{1}{2})}\\
\partial_t \bar {\xi} &=& -\left(\bar {\xi}\partial_x + \bar {\xi}'\right)
A_t^- - \bar {\xi} A_t^{(0)} - \left(\partial_x^2 + J_0 \partial_x
+ J_0' - J_1\right) A_t^{-\frac{1}{2}}\nn
\eqaend
In obtaining (36) we have made use of (35) and the same field redefinition
(11) as in the bosonic case. Note that we have obtained the dynamical
equations in terms of four independent functions $A_t^-$, $A_t^{(0)}$,
$A_t^{-\frac{1}{2}}$ and $ A_t ^{(-\frac{1}{2})}$ and these equations define
the sTB--hierarchy. In order to extract the explicit forms of the equations
we use the same trick as in the bosonic case. The first bi--Hamiltonian
structure and the first set of equations can be obtained by shifting
$J_0(x,t)\rightarrow J_0(x,t) + \lambda$. Again expanding the four
independent functions in powers of $\lambda$ as,
\eqabegin
A_t^-(J_1,J_0,\xi,\bar {\xi},
\lambda) &=& \sum_{j=0}^n A_j(J_1,J_0,\xi,\bar {\xi}) \lambda^{n-j}\nn\\
A_t^{(0)}(J_1,J_0,\xi,\bar {\xi},\lambda) &=& \sum_{j=0}^n B_j
(J_1,J_0,\xi,\bar {\xi}) \lambda^{n-j}\nn\\
A_t^{-\frac{1}{2}}
(J_1,J_0,\xi,\bar {\xi},\lambda) &=& \sum_{j=0}^n \alpha_j
(J_1,J_0,\xi,\bar {\xi}) \lambda^{n-j}\nn\\ A_t^{(-\frac{1}{2})}
(J_1,J_0,\xi,\bar {\xi},\lambda) &=& \sum_{j=0}^n \beta_j
(J_1,J_0,\xi,\bar {\xi}) \lambda^{n-j}\nn
\eqaend
we obtain the following recursion
relations for $j = 0,1,\ldots,n-1$
\eqabegin
-\left(J_1' + 2 J_1 \partial_x\right) A_j + \left(\partial_x^2
- J_0 \partial_x\right) B_j - \left(\xi' + 2 \xi \partial_x\right)
\alpha_j - \bar {\xi} \partial_x \beta_j &=& \partial_x
B_{j+1}\nn\\
-\left(\partial_x^2 + J_0 \partial_x + J_0'\right) A_j
+ 2 \partial_x B_j - \xi \alpha_j + \bar {\xi}
\beta_j &=& \partial_x A_{j+1}\nn\\
-\left(2 \xi \partial_x + \xi'\right) A_j + \xi B_j
- \left(\partial_x^2 - J_1 - J_0\partial_x\right) \beta_j &=&
-\partial_x \beta_{j+1}\\
-\left({\bar \xi}\partial_x + \bar {\xi}'\right)
A_j - \bar {\xi} B_j - \left(\partial_x^2 + J_0 \partial_x
+ J_0' - J_1\right) \alpha_j &=& \partial_x \alpha_{j+1}\nn
\eqaend
and the following dynamical equations for $j=n$,
\eqabegin
\partial_t J_1 &=& -\left(J_1' + 2 J_1 \partial_x\right) A_n +
\left(\partial_x^2
- J_0 \partial_x\right) B_n - \left(\xi' + 2 \xi \partial_x\right)
\alpha_n - \bar {\xi} \partial_x \beta_n\nn\\
\partial_t J_0 &=& -\left(\partial_x^2 + J_0 \partial_x + J_0'\right) A_n
+ 2 \partial_x B_n - \xi \alpha_n + \bar {\xi} \beta_n\nn\\
\partial_t \xi &=& -\left(2 \xi \partial_x + \xi'\right) A_n + \xi B_n
- \left(\partial_x^2 - J_1 - J_0\partial_x\right) \beta_n\\
\partial_t \bar {\xi} &=& -\left(\bar {\xi}\partial_x + \bar {\xi}'\right)
A_n - \bar {\xi} B_n - \left(\partial_x^2 + J_0 \partial_x
+ J_0' - J_1\right) \alpha_n\nn
\eqaend
{}From (37) it is clear that $A_0$, $B_0$, $\alpha_0$ and $\beta_0$ are
constants
independent of $x$. We, therefore, choose $A_0\,=\,-1$, $B_0\,=\,\alpha_0\,=\,
\beta_0\,=\,0$ and obtain from (38) the zeroth order equations as,
\eqabegin
\partial_t J_1 &=& J_1'\nn\\
\partial_t J_0 &=& J_0'\nn\\
\partial_t \xi &=& \xi'\\
\partial_t \bar {\xi} &=& \bar {\xi}'\nn
\eqaend
This is the super extension of chiral wave equation obtained in (17).
Substituting the above zero mode values in the recursion relation (37) we
obtain $A_1\,=\,J_0(x,t)$, $B_1\,=\,J_1(x,t)$, $\alpha_1\,=\,\bar {\xi}
(x,t)$ and
$\beta_1\,=\,-\xi(x,t)$, So, the first order dynamical equations as calculated
from (38) have the form
\eqabegin
\partial_t J_1 &=& \left(J_1' - 2 J_1 J_0 - 2 \xi \bar {\xi}\right)'\nn\\
\partial_t J_0 &=& \left(2 J_1 - J_0' - J_0^2\right)'\nn\\
\partial_t \xi &=& \left(\xi' - 2 J_0 \xi\right)'\\
\partial_t \bar {\xi} &=& \left(-\bar {\xi}' - 2 J_0 \bar {\xi}\right)'\nn
\eqaend
We recognize these equations as the sTB equation [15]. Thus the
whole hierarchy
associated with the sTB equation can be generated from the recursion relations
(37) and the dynamical equations (38). We would also like to point out here
that the sTB fermions $\xi$ and $\bar {\xi}$ do not have specific scaling
dimensions as can be seen from (40). From the first equation of (40) we note
that their sum would have to be 3. So, they could be either $\frac{3}{2}$ each
like $N=2$ sKdV fermions or they could be 2 and 1 like the fermionic fields of
topological field theories. It should be mentioned here that sTB equation (40)
can be reduced to one of the lower equations of $N=2$ sKdV--hierarchy
(eq.(2.33) of ref.[20]) by the following redefinition and rescaling of
fields, $\partial_t
\rightarrow -i\partial_t$, $\partial_x\rightarrow\partial_x$, $J_1\rightarrow
(u+i\phi')$, $J_0\rightarrow 2i\phi$, $\xi\rightarrow (\xi_1 + i\xi_2)$
and $\bar{\xi}\rightarrow (\xi_1 - i\xi_2)$. The next equation of the
sTB--hierarchy with the same redefinitions reduce to $a=4$, $N=2$ sKdV
equation [20,21].
Thus sTB--hierarchy can be thought of as a twisted $N=2$ sKdV--hierarchy.

Now identifying $A_n\,\equiv\,\frac{\delta H_n}{\delta J_1}$, $B_n\,
\equiv\,\frac{\delta H_n}{\delta J_0}$,  $\alpha_n\,\equiv\,
\frac{\delta H_n}{\delta \xi}$ and $\beta_n\,\equiv\,
\frac{\delta H_n}{\delta \bar {\xi}}$, the two Hamiltonian structures can be
easily read out from (37) to be
\eqabegin
\{J_1(x),\, J_0(y)\}_1 &=& \partial_x \delta (x-y)\nn\\
\{\xi(x),\, \bar {\xi}(y)\}_1 &=& -\partial_x \delta (x-y)
\eqaend
and
\eqabegin
\{J_1(x),\, J_1(y)\}_2 &=& -\left[J_1'(x) + 2 J_1(x)\partial_x \right]
\delta (x-y)\nn\\
\{J_1(x),\, J_0(y)\}_2 &=& \left[\partial_x^2 - J_0(x)\partial_x \right]
\delta (x-y)\nn\\
\{J_1(x),\, \xi(y)\}_2 &=& -\left[\xi'(x) + 2 \xi(x)\partial_x \right]
\delta (x-y)\nn\\
\{J_1(x),\, \bar {\xi}(y)\}_2 &=& -\bar {\xi}(x)\partial_x \delta (x-y)\nn\\
\{J_0(x),\, \xi(y)\}_2 &=& -\xi(x)\delta (x-y)\\
\{J_0(x),\, \bar {\xi}(y)\}_2 &=& \bar {\xi}(x)\delta (x-y)\nn\\
\{J_0(x),\, J_0(y)\}_2 &=& 2 \partial_x \delta (x-y)\nn\\
\{\xi(x),\, \bar {\xi}(y)\}_2 &=& -\left[\partial_x^2 - J_0(x)\partial_x
- J_1(x) \right]\delta (x-y)\nn
\eqaend
The second Hamiltonian structure (42) is nothing but the twisted $N=2$
superconformal algebra or topological algebra where $\xi$ and $\bar {\xi}$ have
conformal dimensions 2 and 1 respectively. We would like to comment that the
twisted $N=2$ superconformal algebra [8] contains a finite subalgebra
consisting of
four bosonic and four fermionic generators much like
$OSp(2|2)$ algebra but is not isomorphic to it and may be called `twisted'
$OSp(2|2)$ algebra. It would have been more natural to expect the
sTB--hierarchy to originate from the zero curvature condition on `twisted'
$OSp(2|2)$ rather than $OSp(2|2)$ itself. We have not been able to find any
suitable gauge fixing which will generate sTB--hierarchy this way.

As in the bosonic case the second
bi--Hamiltonian structure can be obtained by shifting $J_1(x,t)\rightarrow
J_1(x,t) + \lambda^2$. Expanding the functions as,
\eqabegin
A_t^-(J_1,J_0,\xi,\bar {\xi},
\lambda^2) &=& \sum_{j=0}^n\, A_j(J_1,J_0,\xi,\bar {\xi}) (\lambda^2)^{n-j}
\nn\\
A_t^{(0)}(J_1,J_0,\xi,\bar {\xi},\lambda^2) &=& \sum_{j=0}^n\, B_j
(J_1,J_0,\xi,\bar {\xi}) (\lambda^2)^{n-j}\nn\\
A_t^{-\frac{1}{2}}
(J_1,J_0,\xi,\bar {\xi},\lambda^2) &=& \sum_{j=0}^n\, \alpha_j
(J_1,J_0,\xi,\bar {\xi}) (\lambda^2)^{n-j}\nn\\
A_t^{(-\frac{1}{2})}
(J_1,J_0,\xi,\bar {\xi},\lambda^2) &=& \sum_{j=0}^n\, \beta_j
(J_1,J_0,\xi,\bar {\xi}) (\lambda^2)^{n-j}\nn
\eqaend
we obtain the following recursion
relations from the hierarchy equations (36) for $j=0,1,\ldots,n-1$
\eqabegin
-\left(J_1' + 2 J_1 \partial_x\right) A_j + \left(\partial_x^2
- J_0 \partial_x\right) B_j - \left(\xi' + 2 \xi \partial_x\right)
\alpha_j - \bar {\xi} \partial_x \beta_j &=& 2\partial_x A_{j+1}\nn\\
-\left(\partial_x^2 + J_0 \partial_x + J_0'\right) A_j
+ 2 \partial_x B_j - \xi \alpha_j + \bar {\xi}
\beta_j &=& 0\nn\\
-\left(2 \xi \partial_x + \xi'\right) A_j + \xi B_j
- \left(\partial_x^2 - J_1 - J_0\partial_x\right) \beta_j &=&
-\beta_{j+1}\\
-\left({\bar \xi}\partial_x + \bar {\xi}'\right)
A_j - \bar {\xi} B_j - \left(\partial_x^2 + J_0 \partial_x
+ J_0' - J_1\right) \alpha_j &=& -\alpha_{j+1}\nn
\eqaend
and for $j=n$ we get the same dynamical equations as in (38). We note from (43)
that $A_0$ is a constant and $\alpha_0\,=\,\beta_0\,=\,0$. $B_j$'s will get
fixed from the second equation in (43). Choosing $A_0\,=\,-1$, we obtain $B_0
\,=\,-\frac{1}{2}J_0(x,t)$ and therefore the zeroth order equation in this case
takes the form
\eqabegin
\partial_t J_1 &=& J_1' - \frac{1}{2} J_0'' + \frac{1}{2} J_0J_0'\nn\\
\partial_t J_0 &=& 0\nn\\
\partial_t \xi &=& \xi' - \frac{1}{2} \xi J_0\\
\partial_t \bar {\xi} &=& \bar {\xi}' + \frac{1}{2} \bar {\xi} J_0\nn
\eqaend
Using the recursion relation (43) we determine $A_1\,=\,\frac{1}{2} J_1 -
\frac{1}{4} J_0' + \frac{1}{8} J_0^2$, $\alpha_1\,=\,-\bar {\xi}' -
\frac{1}{2} \bar {\xi}J_0$, $\beta_1\,=\, -\xi' + \frac{1}{2} \xi J_0$
and $B_1\,=\, \frac{1}{4} J_1' - \frac{1}{8} J_0'' + \frac{1}{4} J_1 J_0 +
\frac{1}{16} J_0^3 + \frac{1}{2} \bar {\xi}\xi$. The first order equations
can be determined from the dynamical equations (38) to have the form
\eqabegin
\partial_t J_1 &=& \frac{1}{4} J_1''' - \frac{3}{2} J_1 J_1' + \frac{3}{2} \xi
\bar {\xi}'' - \frac{3}{2} \xi'' \bar {\xi} + \frac{3}{4} J_1' J_0' -
\frac{3}{8} J_1' J_0^2 + \frac{3}{4} J_1 J_0''\nn\\
& & -\frac{3}{4} J_1 J_0 J_0' + \frac{1}{8} J_0 J_0''' - \frac{3}{16}J_0^3
J_0' + \frac{3}{16}J_0^2 J_0'' + \frac{3}{8}J_0(J_0')^2 - \frac{1}{8}
J_0'''' + \frac{3}{2} (J_0 \xi \bar {\xi})'\nn\\
\partial_t J_0 &=& 0\nn\\
\partial_t \xi &=& \xi''' - \frac{3}{4} J_1' \xi - \frac{3}{2} J_1 \xi' -
\frac{1}{8} J_0'' \xi - \frac{3}{2} J_0 \xi'' - \frac{3}{4} J_0' \xi'
+ \frac{3}{4} J_1 J_0 \xi\nn\\
& & + \frac{3}{8} J_0^2 \xi' + \frac{1}{16} J_0^3 \xi\\
\partial_t \bar{\xi} &=& \bar{\xi}''' - \frac{3}{4} J_1' \bar{\xi} -
\frac{3}{2} J_1 \bar{\xi}' +
\frac{7}{8} J_0'' \bar{\xi} + \frac{3}{2} J_0 \bar{\xi}'' + \frac{9}{4}
J_0' \bar{\xi}' - \frac{3}{4} J_1 J_0 \bar{\xi}\nn\\
& & + \frac{3}{8} J_0^2 \bar{\xi}' - \frac{1}{16} J_0^3 \bar{\xi} +
\frac{3}{4} J_0J_0'\bar{\xi}\nn
\eqaend
Eq.(45) resembles again as one of the $N=2$ sKdV equation
(Eq.(2.27) of ref.[20])
which is not supersymmetric. One can check indeed that they are identical with
the redefinitions, $\partial_t\rightarrow -\frac{1}{4}\partial_t$, $\partial_x
\rightarrow \partial_x$, $J_0\rightarrow 2i\phi$, $J_1\rightarrow u+i\phi'$,
$\xi\rightarrow\frac{1}{\sqrt {2}}(\xi_1+\xi_2)$, $\bar{\xi}\rightarrow
\frac{1}{\sqrt {2}}(\xi_1-\xi_2)$. The first
Hamiltonian structure in this case can be read out from (43) to be,
\eqabegin
\{J_1(x),\,J_1(y)\}_1 &=& 2\partial_x \delta (x-y)\nn\\
\{\xi(x),\,\bar{\xi}(y)\}_1 &=& -\delta (x-y)
\eqaend
and the second Hamiltonian structure remains the same as in (42). So, like
in the bosonic case, the zero curvature formulation leads to two
bi--Hamiltonian systems.

We have seen that $N=2$ sKdV--hierarchy can be obtained from
the sTB--hierarchy through some redefinition of fields, but this is quite
expected since the Hamiltonian structures are related to each other by
replacing $J_1\rightarrow J_1 + \frac{1}{2}J_0'$.
But, we would like to emphasize
that the choice of gauge fixing eq.(34) is not at all obvious as compared to
the gauge fixing in $N=2$ sKdV--hierarchy [20] and further, we had to make a
nonlinear field redefinition eq.(11) to recast the hierarchy in the proper
form. Our procedure also clarifies how other supersymmetric integrable systems
which are known to be contained in sTB--hierarchy (for example, supersymmetric
Non--Linear Schrodinder equation [22]) can be obtained in the zero curvature
approach.
Finally, in the following we would like to make an observation on the
topological symmetry of sTB--hierarchy. It can be readily verified that the
sTB--hierarchy equation (36) is invariant under the following two sets of
supersymmetry transformations (note that these symmetries respect
only one of the hierarchies obtained by $\phi\rightarrow\phi+\lambda$),
\eqabegin
\bar {\delta} J_1 &=& 0\,\,\,\,\,\,\,\,\,\,\,\,\,\,\,\,\,\,\,\,\,\,
\,\,\,\,\,\,\,\,\,\,\bar {\delta}
A_t^{(-\frac{1}{2})}\,\, =\,\, -\bar{\epsilon} A_t^{(0)}\nn\\
\bar {\delta}\phi &=& -\bar {\epsilon} \bar {\xi}\,\,\,\,\,\,\,\,\,\,\,\,\,\,
\,\,\,\,\,\,\,\,\,\,\,\,\,
\bar {\delta}A_t^- \,\,=\,\, -\bar
{\epsilon} A_t^{-\frac{1}{2}}\nn\\
\bar{\delta}\bar{\xi} &=& 0\,\,\,\,\,\,\,\,\,\,\,\,\,\,\,\,\,\,\,\,\,\,\,\,
\,\,\,\,\,\,\,\,\,\,\bar {\delta}
A_t^{(0)} \,\,=\,\, 0\\
\bar{\delta}\xi &=& \bar{\epsilon} J_1\,\,\,\,\,\,\,\,\,\,\,\,\,\,\,\,\,
\,\,\,\,\,\,\,\,\,\,\bar{\delta}
A_t^{-\frac{1}{2}} \,\,=\,\, 0\nn
\eqaend
and
\eqabegin
\delta J_1 &=& \epsilon \xi'\,\,\,\,\,\,\,\,\,\,\,\,\,\,\,\,\,\,\,\,\,\,\,\,
\,\,\,\,\,\,\,\,\,\,\,\delta A_t^-
\,\,=\,\,-\epsilon A_t^{(-\frac{1}{2})}\nn\\
\delta\phi &=& \epsilon \xi\,\,\,\,\,\,\,\,\,\,\,\,\,\,\,\,\,\,\,\,\,\,\,\,
\,\,\,\,\,\,\,\,\,\,\,\delta
A_t^{(0)}\,\,=\,\, -\epsilon \partial_x A_t^{(-\frac{1}{2})}\nn\\
\delta\bar{\xi} &=& \epsilon\left(J_1 - J_0'\right)\,\,\,\,\,\,\,\,\,\,\,\,
\delta
A_t^{-\frac{1}{2}}\,\,=\,\,\epsilon\left(A_t^{(0)}-\partial_x A_t^-
\right)\\
\delta\xi &=& 0\,\,\,\,\,\,\,\,\,\,\,\,\,\,\,\,\,\,\,\,\,\,\,\,\,\,\,\,\,
\,\,\,\,\,\delta
A_t^{(-\frac{1}{2})}\,\,=\,\,0\nn
\eqaend
We note, first of all, that both these symmetries are of nilpotent type i.e.
$\delta^2\,\,=\,\,\bar{\delta}^2\,\,=\,\,0.$ Also, from the first set of
equation in (47), we notice that both $J_1$ (which can be identified with
energy--momentum tensor) and $\bar{\xi}$ (fermionic current) are
$\bar{\delta}$-exact, the basic ingredients of a topological
field theory [7,8].
In fact, from the Hamiltonian structure (42), we can read out the two symmetry
charges of (47) and (48) to be $\int\, dx\,\bar{\xi}(x)$ and
$\int\, dx\,\xi(x)$, which have conformal weights zero and one respectively
much like the topological $Q$ and $G$-charges of a topological field
theory.

To conclude, we have shown how TB--hierarchy can be obtained from the
zero curvature condition associated with the gauge group
$SL(2,R)\otimes U(1)$. sTB--hierarchy has similarly been shown to follow from
the zero curvature condition of the gauge supergroup $OSp(2|2)$. It is shown
that both these integrable systems contain two bi--Hamiltonian structures.
We have also made the field identifications by which sTB--hierarchy can be
converted to $N=2$ sKdV--hierarchy and thus pointed out that
sTB--hierarchy is nothing but the twisted $N=2$ sKdV--hierarchy. Topological
algebras arise naturally as the second Hamiltonain structure in our zero
curvature formulation of both TB and sTB--hierarchy indicating a close
relationship of these integrable models with 2d topological theories.
\parbigskipn
\noindent{\bf ACKNOWLEDGEMENTS:}
\parmedskip
We would like to thank J. C. Brunelli for discussions as well as for
collaboration in the earlier stages of the work. One of us (S.R.) also
thanks W. -J. Huang for discussions at an early stage of this work and
J. Mas for interesting comments. The work
of S.R. is supported in part by the Spanish Ministry of Education (MEC)
fellowship. This work was supported in part by the U.S. Department of Energy
grant DE-FG02-91ER40685.
\parbigskipn
\noindent{\bf REFERENCES:}
\parmedskip
\begin{enumerate}
\item M. A. Olshanetsky and A. M. Perelomov, Phys. Rep. 71 (1980) 315.
\item L. D. Fadeev and L. A. Takhtajan, {\it Hamiltonian Methods in the Theory
of Solitons}, (Springer--Verlag, 1987).
\item A. Das, {\it Integrable Models}, (World Scientific, Singapore, 1989).
\item L. A. Dickey, {\it Soliton Equations and Hamiltonian Systems},
(World Scientific, Singapore, 1991).
\item J. L. Gervais and A. Neveu, \np 209 (1982) 125.
\item D. Gross and A. Migdal, \np 340 (1990) 333; E. Brezin and V. Kazakov,
\pl 236 (1990) 144; M. R. Douglas, \pl 238 (1990) 176.
\item E. Witten, \cmp 117 (1988) 353; \cmp 118 (1988) 411; T. Eguchi and S. K.
Yang, Mod. Phys. Lett. A4 (1990) 1693; R. Dijkgraaf and E. Witten, \np 342
(1990) 486.
\item R. Dijkgraaf, E. Verlinde and H. Verlinde, \np 348 (1991) 435; Preprint
PUPT-1217/IASSNS-HEP-90/80.
\item E. Witten, Surv. Diff. Geom. 1 (1991) 243.
\item M. Kontsevich, Peprints MPI/91-47, MPI/91-77.
\item B. A. Kupershmidt, \cmp 99 (1985) 51.
\item L. Bonora and C. S. Xiong, \pl 285 (1992) 191; Int. Jour. Mod. Phys.
A8 (1993) 2973; H. Aratyn, L. A. Ferreira, J. F. Gomes and A. H. Zimerman,
\np 402 (1993) 85.
\item L. J. F. Broer, Appl. Sci. Res. 31 (1975) 377.
\item D. J. Kaup, Progr. Theor. Phys. 54 (1975) 396.
\item J. C. Brunelli and A. Das, \pl 337 (1994) 303.
\item J. C. Brunelli and A. Das, \pl 354 (1995) 307;
Int. J. Mod. Phys.  A10 (1995) 4563.
\item S. S. Chern and C. K. Peng, Manus. Math. 28 (1975) 145.
\item S. Okubo and A. Das, \pl 209 (1988) 31; A. Das and S. Okubo, Ann. Phys.
190 (1989) 215.
\item A. Das, W. -J. Huang and S. Roy, Int. Jour. Mod. Phys. A7 (1992) 3447.
\item A. Das, W. -J. Huang and S. Roy, Int. Jour. Mod. Phys. A7 (1992) 4293.
\item C. A. Laberge and P. Mathieu, \pl 215 (1988) 718.
\item S. Krivonos and A. Sorin, Preprint JINR E2-95-172, hep-th/9504084.
\end{enumerate}
\vfil

\eject
\end{document}